\documentclass{jpsj-suppl}

\title{Strangeness At Extremes}

\author{Laura \textsc{Tolos}$^{1,2}$, Daniel \textsc{Cabrera}$^{2}$, Kanchan \textsc{Khemchandani}$^{3}$ , Alberto \textsc{Martinez-Torres}$^{4}$,
Elena \textsc{Bratkovskaya}$^{2}$, Joerg \textsc{Aichelin}$^{5}$, Marina \textsc{Nielsen}$^{4}$ and Fernando S~\textsc{Navarra}$^{4}$ }

\inst{$^{1}$Instituto de Ciencias del Espacio (IEEC/CSIC), Campus UAB, 08193 Cerdanyola del Valles, Spain.\\
$^{2}$ Frankfurt Institute for Advanced Studies and Institut f\"ur Theoretische Physik, J. W. Goethe-Universit\"at,  60438 Frankfurt (M), Germany\\
$^{3}$ Faculdade de Tecnologia, Universidade do Estado do Rio de Janeiro, 27537-000, Resende, RJ, Brazil \\
$^{4}$  Instituto de Fisica, Universidade de Sao Paulo, 05314-970 Sao Paulo, SP, Brazil\\
$^{5}$ Subatech,  IN2P3/CNRS, \'Ecole de Mines de Nantes, Universit\'e de Nantes, France}

\email{tolos@ice.csic.es}

\recdate{}

\abst{We study the properties of strange mesons in vacuum and in the hot nuclear medium within unitarized coupled-channel effective theories. We determine transition probabilities, cross sections and scattering lengths for strange mesons. These scattering observables are of fundamental importance for understanding the dynamics of strangeness production and propagation in heavy-ion collisions. }

\kword{strange meson, coupled-channel unitarized theory,  heavy-ion collisions}

\begin{document}
\maketitle

\section{Introduction}

Strangeness in hot and dense matter has been matter of intense investigation over the past years in connection with the study of the composition of neutron stars \cite{Kaplan:1986yq}, the properties of exotic atoms \cite{Friedman:2007zza}, and strangeness production in heavy-ion collisions (HICs) \cite{Fuchs:2005zg,Forster:2007qk,Hartnack:2011cn}. In particular, understanding the dynamics of light strange mesons, such as $\bar K$, $\bar K^*$, $K$ or $K^*$ mesons, in vacuum and in the nuclear environment is still a challenge for theoretical models. 

The production of $K$ and $\bar K$ close to the threshold energy has been thoroughly investigated in HICs at SIS energies \cite{Fuchs:2005zg,Forster:2007qk,Hartnack:2011cn,Cassing:2003vz,Tolos:2003qj}. The analysis of experimental data together with microscopic transport approaches have allowed to draw several conclusions regarding the production mechanisms  and the freeze-out conditions of strange mesons. Still, a simultaneous description of all observables involving $\bar K$ production is missing \cite{Hartnack:2011cn}.  Also, recent results by the HADES Collaboration on $K^0$ production have been reported in $p+p$  and $p+Nb$ collisions at 3.5 GeV \cite{Agakishiev:2014nim,Agakishiev:2014moo}, whereas a deep sub-threshold $K^{*0}$ production has been reported in Ar+KCl collisions \cite{hades}. The importance of the hadronic interactions  has been also realized for the $K^{*0}$ production in Au+Au and Cu+Cu at $\sqrt{s_{NN}}=$62.4 and 200 GeV collisions by the STAR Collaboration \cite{star}, while attenuation of the $K^*$ and $\bar K^*$ states in the hadronic phase of the expanding fireball in HICs has been observed by the NA49 Collaboration \cite{NA49}. Such findings highlight the importance of a reliable determination of the interaction of strange mesons with the surrounding nuclear environment.

In this paper we aim at investigating the properties of  strange mesons in vacuum and hot nuclear matter, so as to address the possible consequences for strangeness production and propagation in heavy-ion collisions.

\section{$\bar K$ mesons in hot nuclear matter}

Strange pseudoscalar mesons in nuclear matter at finite temperature have been studied within a self-consistent coupled-channel approach based on the SU(3) meson-baryon chiral Lagrangian (see \cite{Tolos:2006ny,Tolos:2008di} and references therein). Within this approach, the $\bar K$ meson interaction in nuclear matter has been analyzed, finding a shallow attractive $\bar K$ potential at normal matter density ($\rho_0=0.17~{\rm fm}^{-3}$) and a broad spectral function \cite{Tolos:2006ny,Tolos:2008di}.  Recently we have improved on this model by implementing the unitarization of $\bar K N$ scattering amplitudes in both $s$- and $p$- waves at finite density and temperature \cite{Cabrera:2014lca}.
In this manner, we have access to the off-shell in-medium coupled-channel scattering amplitudes. For example, for $K^- p$ scattering, the coupled channels comprise $K^- p$, $\bar{K}^0n$, $\pi^0
\Lambda$, $\pi^0 \Sigma^0$, $\eta \Lambda$, $\eta \Sigma^0$,  $\pi^+
\Sigma^-$, $\pi^- \Sigma^+$, $K^+ \Xi^-$ and $K^0 \Xi^0$; whereas for $K^- n$ scattering we have $K^- n$, $\pi^0 \Sigma^-$, $\pi^- \Sigma^0$, $\pi^- \Lambda$, $\eta \Sigma^-$ and $K^0 \Xi^-$.

\begin{figure}[t]
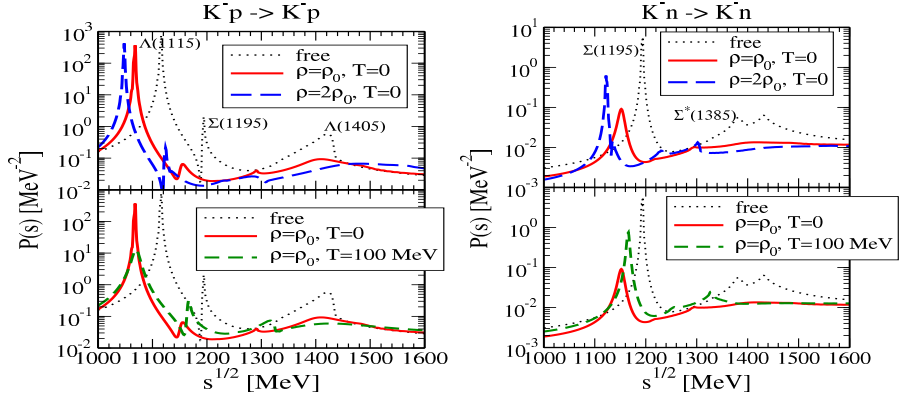

\centering
\includegraphics[height=5.2cm,width=0.38\textwidth,clip]{fig1a.eps} 
\includegraphics[height=5.2cm,width=0.38\textwidth,clip]{fig1b.eps} 
\caption{In-medium transition probability ${\cal P}$ at zero total three-momentum of the meson-baryon pair for $K^-p$ (left) and $K^- n$ (right) elastic reactions. All plots are taken from Ref.~\cite{Cabrera:2014lca}.}
\label{fig:kmp-vs-kmn}       
\end{figure}

The transition probability for a given reaction, ${\cal P}$, is defined as the angular integrated modulus-squared scattering amplitude (including all partial waves, $L=0,1$), averaged over the total angular momentum ($S=1/2, 3/2$). The cross section for the process $i\to j$ then follows as $\sigma_{ij} = \frac{1}{8 \pi} \frac{M_i M_j}{s} \frac{\tilde{q}_j}{\tilde{q}_i}\,{\cal P}_{ij}$, with $\tilde{q}_i$ the center-of-mass three-momentum and $M_i$ the baryon mass in channel $i$. The total cross section for a given reaction  is then given by the sum over the partial cross sections for all  coupled channels, $\sigma^i_{\rm tot} = \sum_j \sigma_{ij}$. 

In  Fig.~\ref{fig:kmp-vs-kmn} we show the transition probability for the $K^-p$ and $K^- n$ elastic reactions as a function of the meson-baryon center-of-mass energy at total vanishing three-momentum. The $K^-p$ state is an admixture of isospin $I=0,1$ and, therefore, the two $\Lambda$ resonances and the $\Sigma(1195)$ show up in the spectrum, as seen in  the left panel of Fig.~\ref{fig:kmp-vs-kmn}. The $K^-n$ reaction is pure $I=1$ and consequently only the isovector hyperon excitations are present in the right panel of Fig.~\ref{fig:kmp-vs-kmn}. This explains the dramatic difference between the $K^-p$ and $K^- n$ cross sections in vacuum. In the medium,  the structure of the $\Lambda(1405)$ is washed out at normal matter density. The $p$-wave ground states experience moderately attractive mass shifts as density is increased, whereas the $\Sigma^*(1385)$ is largely broadened due to the opening of in-medium decay channels such as $\Sigma^*\to\Lambda NN^{-1}$, $\Sigma NN^{-1}$. The effect of temperature is particularly appreciable as a broadening of the $\Lambda(1115)$, $\Sigma(1195)$ and $\Sigma^*(1385)$ as compared to the vacuum case.

A more quantitative analysis of the transition amplitudes allows us to extract information on the single-particle  potentials of $\Lambda(1115)$, $\Sigma(1195)$ and $\Sigma^*(1385)$ hyperons at finite momentum, density and temperature \cite{Cabrera:2014lca}. At finite nuclear densities both the $\Lambda$ and the $\Sigma$ experience an attractive potential of roughly -50 and -40~MeV, respectively, at normal matter density and zero temperature. Both hyperons acquire a finite decay width, reflecting the probability to be absorbed by the nuclear medium or have quasi-elastic scattering processes at finite density and temperature. The $\Sigma^*$ develops a much smaller attractive potential of about -10~MeV at $\rho=\rho_0$ and zero temperature, that turns into a small repulsion for increasing densities. Its decay width is notably enhanced at finite density due to the opening of new absorption channels as pions are dressed. The effect of the temperature in this case is moderate due to the important phase space already available at zero temperature. Also, we find that the optical potentials of the $\Lambda$, $\Sigma$ and $\Sigma^*$ have a smooth behavior  with momentum \cite{Cabrera:2014lca}.

All the results presented here are complementary to the ones obtained in our previous work \cite{Tolos:2008di}, where the $\bar K$ spectral function and nuclear optical potential were provided. Altogether, they allow for a systematic analysis of medium effects in the strangeness -1 sector within transport approaches such as PHSD and IQMD \cite{Hartnack:2011cn}.

\section{$KN$ and $K^*N$ interactions in free space}

\begin{figure}[t]
\includegraphics[height=4.5cm,width=0.31 \textwidth]{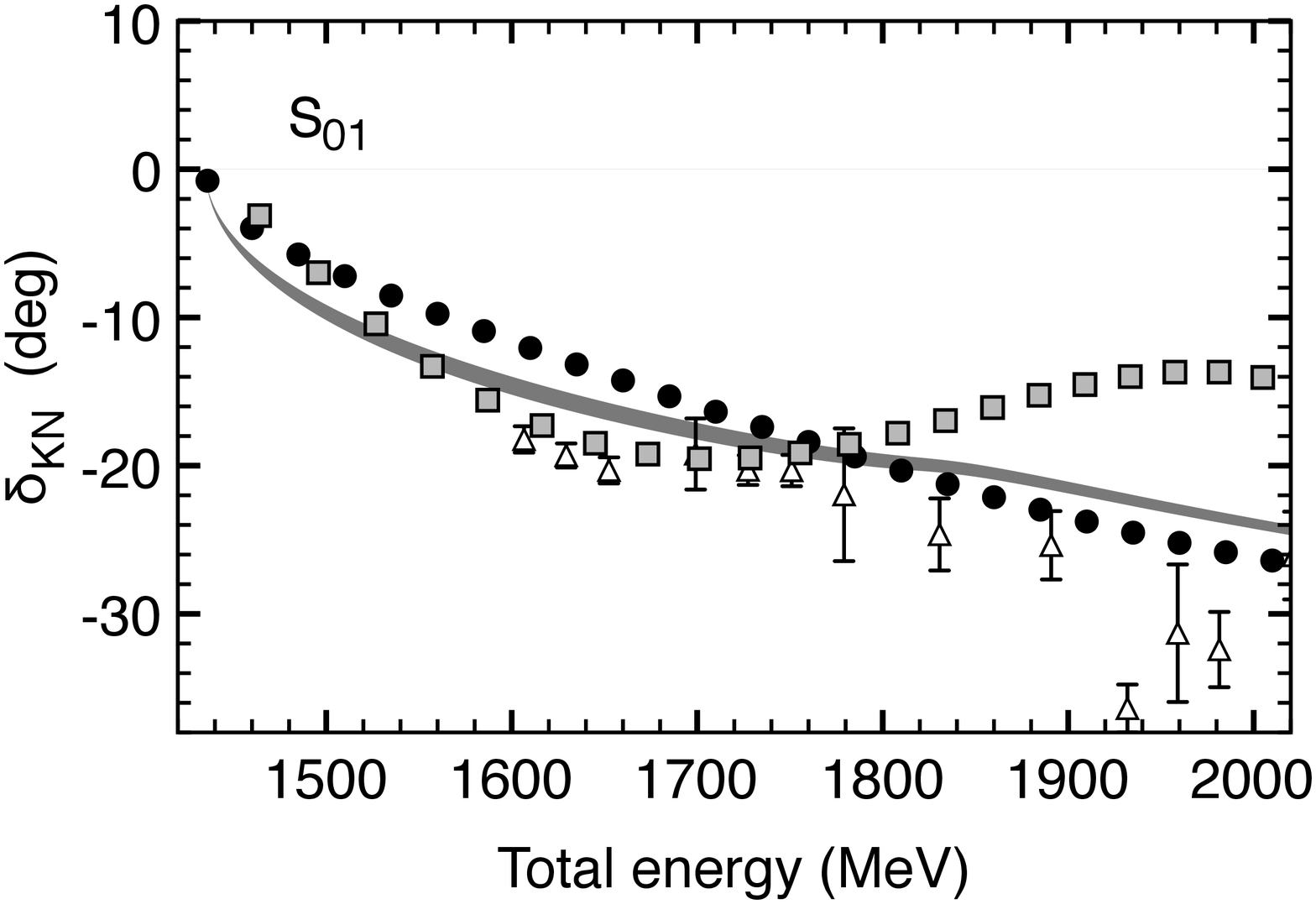}
\includegraphics[height=4.5cm, width=0.31\textwidth]{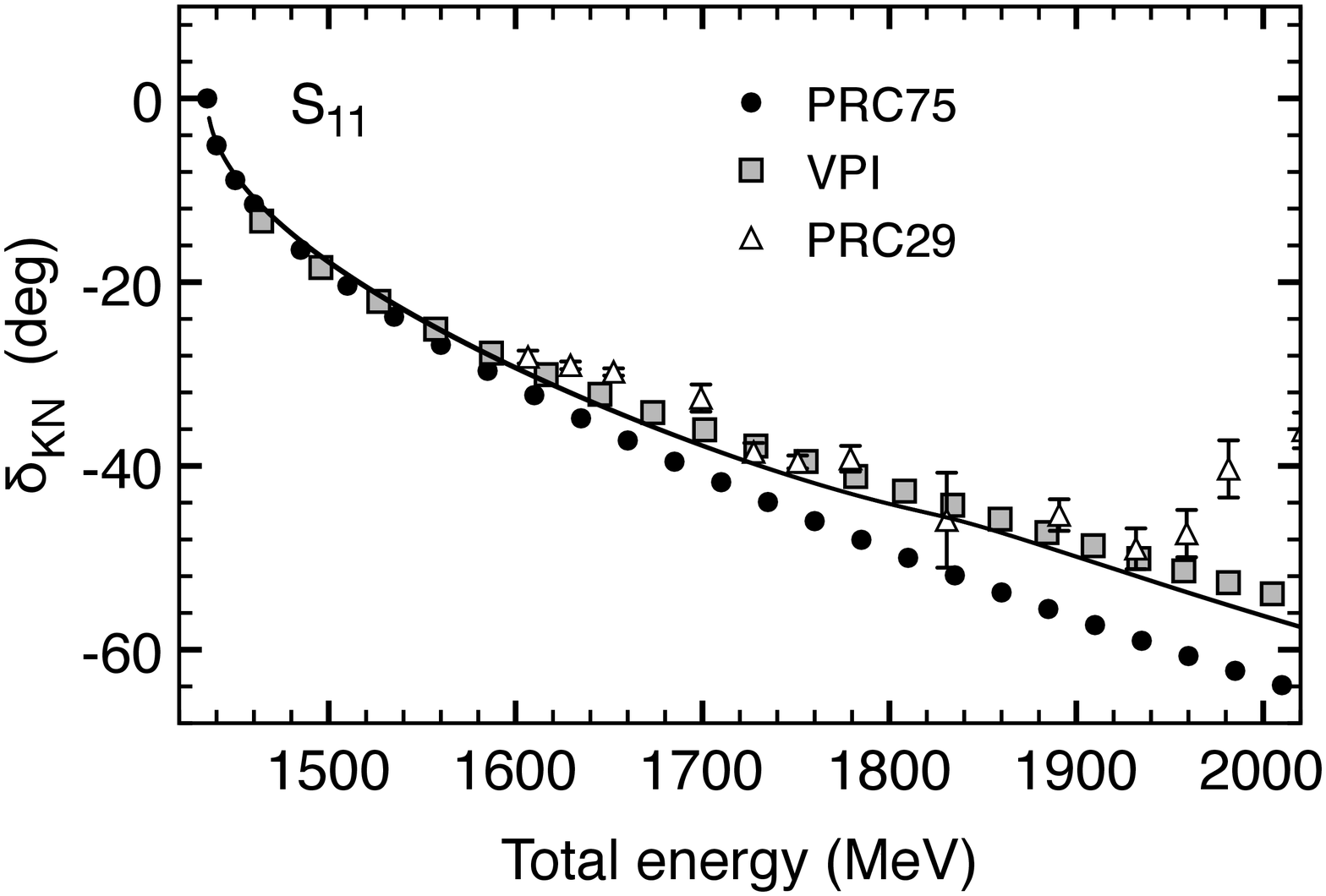}
\includegraphics[height=4.5cm, width=0.3\textwidth]{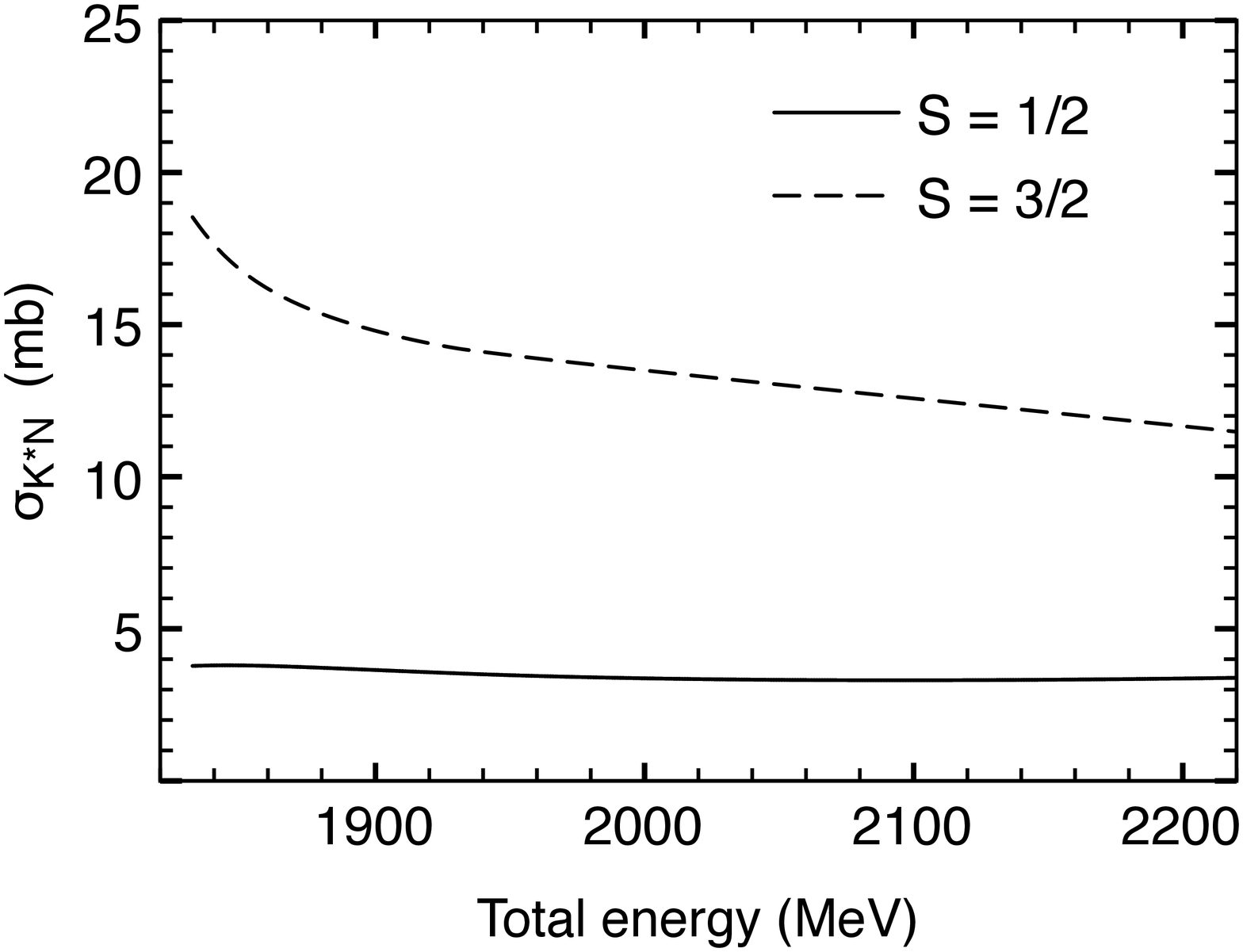}
\caption{Left and middle panels: Scattering phase shifts for the $KN$ system.  The data from the partial wave analysis  are taken from Refs.~\cite{prc75,gwu,prc29}. Right panel: Total cross sections  for $K^*N$ in the different spin channels. All plots are taken from Ref.~\cite{Khemchandani:2014ria}. \label{fig:kn}}
\end{figure}

In this section we present a study for $KN$ and $K^* N$ using a formalism based on the chiral Lagrangian and the theory of hidden local symmetry (HLS) of Ref.~\cite{bando}. The amplitudes for vector meson-baryon channels are obtained from the $s$-, $t$-, $u$- channel diagrams  and a contact interaction, all derived  from Lagrangians based on HLS. The pseudoscalar meson-baryon interactions are calculated by relying on the Weinberg-Tomozawa theorem. The transition amplitudes between the systems consisting of pseudoscalars and vector mesons are computed  by extending the Kroll-Ruderman term for pion photoproduction replacing the photon by a vector meson.  In addition, the exchange of light hyperon resonances, such as $\Lambda(1405)$ and $\Lambda(1670)$ in the $u$-channel, is included  for the $KN \leftrightarrow KN$,  $K^*N \leftrightarrow K^*N$ and $K^*N \leftrightarrow KN$ processes \cite{Khemchandani:2014ria}.

With all these ingredients, we construct the $s$-wave $KN$ and $K^*N$ interactions and solve the Bethe-Salpeter equation in its on-shell factorization form. The subtraction constants required to calculate the loops of the Bethe-Salpeter equation are obtained by fitting our $KN$ amplitudes to the data available for the isospin 0 and 1 $s$-wave phase shifts, separately. The best fit is obtained with the values of the subtraction constants: $b^{I=0}_{KN}=-6.82$, $b^{I=0}_{K^*N}=1.84$, $b^{I=1}_{KN}=-1.59$ and $b^{I=1}_{K^*N}=-1$ for the regularization scale $\mu$ fixed to 630 MeV. 

The $s$-wave phase shifts are shown in the left and middle panels of Fig. 2, where we vary the value of the form-factor cutoff  from the exchange of light hyperon resonances in the $u$-channel between 650 MeV$-$1000 MeV \cite{Khemchandani:2014ria}. It can be seen that there is a good agreement between our results and the partial wave analysis data of Refs.~\cite{prc75,gwu,prc29} for both isospins.  The results for the phase shifts in isospin 0 are more sensitive to the form-factor cutoff than the isospin 1 ones due to a weak $KN \leftrightarrow K^*N$ amplitude in the isospin 1 case.

Furthermore, we can calculate the scattering lengths, $a^{I,S}$, at threshold for the $KN$ system in the different isospin sectors  $I$ and $S=1/2$. We find $a^{0,1/2}_{KN} =-0.16$ fm and $a^{1,1/2}_{KN} =-0.29$ fm. The values found by different partial wave analysis groups for the $KN$ scattering lengths range from  $-0.105 \pm 0.01$ fm \cite{prc75} to $-0.23 \pm 0.18$ fm \cite{martin}, for isospin 0, and between $-0.286 \pm 0.06$ fm to $-0.308 \pm 0.003$ fm \cite{prc75}, for the isospin 1 case. Thus, we find that our results are compatible with the available data.

With the assurance of a reasonable agreement between our results and the available data on $KN$, we analyze the $K^*N$ channel. We obtain the scattering lengths as well as the cross sections for different isospin-spin sectors. The scattering lengths are: $a^{0,1/2}_{K^*N} ({\rm fm})= (0.2,0.03)$, $a^{0,3/2}_{K^*N} ({\rm fm})=(-0.08,0.04)$, $a^{1,1/2}_{K^*N} ({\rm fm})=(0.1,0.0)$ and $a^{1,3/2}_{K^*N} ({\rm fm})=(-0.31,0.03)$.  Moreover, in the right panel of Fig.~\ref{fig:kn} we show the $K^*N$ total cross sections for both spins. 

The results on cross sections and scattering lengths for the $KN$ and $K^*N$ systems  are, indeed, of special interest for $K$ and $K^*$ production in $p + p$ and $p + A$ collisions, as reported by HADES \cite{Agakishiev:2014nim,Agakishiev:2014moo,hades}, STAR \cite{star} and  NA49 \cite{NA49} Collaborations.

\section*{Acknowledgments}
This research was supported by Ministerio de Econom\'{\i}a y Competitividad under contract FPA2013-43425-P,  the LOEWE Center "HIC for FAIR" as well as BMBF Grant Nr. 06FY7080 and  Nr. 05P12RFFCQ. L.T.
acknowledges support from the Ram\'on y Cajal Research Programme from Ministerio de Econom\'{\i}a y Competitividad.

\end{document}